\documentclass[prb,reprint,superscriptaddress,longbibliography]{revtex4-1}
\usepackage{graphicx}
\usepackage[caption=false]{subfig}
\usepackage{bm}
\usepackage{amsmath}
\usepackage{amssymb}
\usepackage{mathtools}
\usepackage{siunitx}
\usepackage{multirow}
\usepackage{array}
\usepackage{float}
\usepackage{xcolor}
\usepackage{longtable}
\usepackage[version=4]{mhchem}
\usepackage{hyperref}
\hypersetup{colorlinks=true,
            linkcolor=red,
            citecolor=blue,
            urlcolor=orange}
 
\newcolumntype{K}[1]{>{\centering\arraybackslash}p{#1}}
\setcitestyle{numbers,round}

\AtBeginDocument{\usepackage{booktabs}}

\begin{document}

\title{Titanium-hydrogen interaction at megabar pressure}

\author{Arslan B. Mazitov}
\email{arslan.mazitov@phystech.edu}
\affiliation{Dukhov Research Institute of Automatics (VNIIA), Moscow 127055, Russian Federation}
\affiliation{Moscow Institute of Physics and Technology, 141700, 9 Institutsky lane, Dolgoprudny, Russian Federation}

\author{Artem R. Oganov}

\affiliation{Skolkovo Institute of Science and Technology, Skolkovo Innovation Center 143026, 3 Nobel Street, Moscow, Russian Federation}
\affiliation{Dukhov Research Institute of Automatics (VNIIA), Moscow 127055, Russian Federation}
\affiliation{International Center for Materials Discovery, Northwestern Polytechnical University, Xi'an, 710072, China}

\author{Alexey V. Yanilkin}
\affiliation{Dukhov Research Institute of Automatics (VNIIA), Moscow 127055, Russian Federation}
\affiliation{Moscow Institute of Physics and Technology, 141700, 9 Institutsky Lane, Dolgoprudny, Russian Federation}

\date{\today}   

\begin{abstract}
    The process of transport of metal particles (\textit{ejecta}) in gases is the subject of recent works in the field of nuclear energetics. We studied the process of dissolution of titanium ejecta in warm dense hydrogen at megabar pressure. Thermodynamic and kinetic properties of the process were investigated using classical and quantum molecular dynamics methods. We estimated the dissolution time of ejecta, the saturation limit of titanium atoms with hydrogen and the heat of dissolution. It was found that particles with a radius of 1 $\mu m$ dissolve in hydrogen in time of $1.5 \cdot 10^{-2} \ \mu s$, while the process of mixing can be described by diffusion law. The presented approach demonstrates the final state of the titanium-hydrogen system as a homogenized fluid with completely dissolved titanium particles. This result can be generalized to all external conditions under which titanium and hydrogen are atomic fluids.
\end{abstract}

\maketitle

\section{Introduction \label{sec:intro}}
Investigations of the interaction in metal-hydrogen systems at different external conditions are of both fundamental and practical value. Metals and hydrogen can form hydrides at normal conditions that are promising for hydrogen storage \cite{schlapbach2011hydrogen, SAKINTUNA20071121} as well as moderators, reflectors or shield components for high-temperature mobile nuclear reactors \cite{mueller2013metal}. In 2004, the hypothesis of high-temperature superconductivity of hydrogen-rich hydrides under high pressure was proposed \cite{ashcroft2004}. This was subsequently confirmed both theoretically  \cite{niobium, esfahani2016superconductivity, kruglov2017hs, kruglov2017uh, thorium, kvashnin2017iron, semenok2018actinium} and experimentally \cite{ferrum, drozdov2015conventional, lanthanium}. At pressures of about 100 GPa, stable hydrides with high hydrogen content, such as $\mathrm{NbH_{6}}$ \cite{niobium},  $\mathrm{FeH_{5}}$ \cite{ferrum, kvashnin2017iron}, $\mathrm{LaH_{10}}$ \cite{lanthanium},  $\mathrm{ThH_{10}}$ \cite{thorium}, $\mathrm{UH_{8}}$ \cite{kruglov2017uh}, $\mathrm{SnH_{14}}$ \cite{esfahani2016superconductivity} and  $\mathrm{AcH_{16}}$ \cite{semenok2018actinium} were found to be thermodynamically stable.

The practical interest comes from the processes of inertial thermonuclear fusion \cite{boiko1999fusion}, which are often accompanied by separation of ejecta particles from the interior surface of the fuel target. In \cite{buttler2017ejecta}, it was shown that at their initial stage at the pressure of 3 atm, ejecta with a radius of approximately 1 $\mu m$ can form hydrides within 1 $\mu s$. With further increase of pressure, the saturation degree of the ejecta with hydrogen can significantly increase, and this will negatively affect the amount of pure hydrogen in the system. Finally, recent calculations  \cite{ticknor2016plasma, clerouin2017enhancement} demonstrated that at temperatures of between $10^5-10^7 $ K and pressures of several hundred gigapascals, the metal-hydrogen system behaves as a two-component plasma, while the process of mixing can be described by diffusion law. Therefore, the investigation of the degree and kinetics of saturation in the intermediate region of pressures and temperatures is relevant. 

In this paper, we study the interaction between titanium and hydrogen. In order to estimate the hydrogen content in stable compounds at $\mathrm{P \simeq 100 \ GPa}$, we performed a variable-composition search using the  evolutionary algorithm USPEX \cite{oganov2006crystal, oganov2011evolutionary, lyakhov2013new}. The process of mixing of titanium-hydrogen system at temperatures $\mathrm{T > 3 \cdot 10^3\  K}$ and megabar pressure was evaluated using quantum (QMD) and classical (MD) molecular dynamics methods. The calculations were performed employing VASP (QMD) \cite{kresse1993, kresse1996,kresse1999} and LAMMPS  (MD)  \cite{lammps} codes. Under these conditions, titanium is above the melting curve and hydrogen is dissociated \cite{stutzmann2015, tamblyn2010}, which allows us to consider the components as atomic fluids. 

One of the main issues underlying the MD method is to find the interatomic potential, which would correctly describe the atomic forces and energies of structures. These values can be calculated from first principles (\textit{ab initio}) using density functional theory (DFT)  \cite{hohenberg1964inhomogeneous, kohn1965self}, but this is computationally expensive. For systems with more than $10^3$ atoms it becomes vital to use flexible mathematical expressions with a large number of free parameters based on machine learning (ML) potentials. In practice, the accuracy of these potentials is higher than that of common classical potentials \cite{kruglov2017}. Today, a number of possible options for building ML potentials exist:
\begin{enumerate}
    \item Linear and Gaussian regressions based on the atomic-environment descriptors \cite{bartok2010gaussian, bartok2013, bishop2006, kruglov2017}
    \item Invariant tensor polynomials \cite{shapeev2016moment}
    \item Neural networks \cite{behler2015ml}
\end{enumerate}

Using \textit{ab initio} data on energies and forces of relatively small systems, a ML potential can be \textit{trained} by adjusting internal parameters, and then used to predict the  behavior of large systems.

\section{Methods \label{sec:methods}}

\subsection{Evolutionary search \label{subsec:uspex}}

The evolutionary algorithm USPEX provides a systematic approach to the search for stable structures of compounds at given pressure. Recently, several studies in the field of prediction of crystal structure of hydrides at various pressures were carried out \cite{niobium, esfahani2016superconductivity, kruglov2017hs, kruglov2017uh, thorium, kvashnin2017iron}. We performed a variable-composition search at P = 100 GPa in order to predict composition and crystal structures of thermodynamically stable titanium hydrides at 0 K. The first generation (80 structures) was produced randomly with up to 16 atoms in the primitive unit cell. The next generations of the structures were obtained by applying heredity (40 \%), softmutation (20 \%),transmutation (20 \%) operators. 20 \% were produced using random symmetry and random topology generators. Each generated crystal structure was relaxed using VASP code with PAW pseudopotentials with four valence electrons (3d$^3$4s$^1$) for Ti \cite{kresse1999, PAW}. Exchange-correlation effects were described within the generalized gradient approximation (Perdew-Burke-Ernzerhof functional) \cite{PBE}. The energy cutoff of plane waves was set to 400 eV, and the Brillouin zone was sampled with k-points grid, centered at $\Gamma$ (0,0,0) point with a resolution of $2 \pi \cdot 0.05$ \AA$^{-1}$.

\subsection{QMD calculation \label{subsec:QMD}}

QMD simulations were carried out to investigate the mixing process based on first principles. In this way, the pure components of titanium (48 atoms) and hydrogen (256 atoms) at P $\simeq$ 100 GPa and T $\simeq$ 3000 K were prepared by preliminary relaxation. Thereafter, the MD of each system during $2 \cdot 10^3$ steps (each of 0.2 fs) was performed. The radial distribution function (RDF) of the final structures confirmed that titanium is in the liquid state and hydrogen is fluid. Finally, these two systems with pure components were placed close to each other to start mixing. The simulation of mixing was carried out during $2 \cdot 10^3$ steps in the NVE ensemble so the average temperature could change during mixing.

\subsection{Classical MD calculation \label{subsec:MD}}
Although mixing in the Ti-H system can be qualitatively investigated with a small system, quantitative analysis requires systems much larger than those discussed in the previous section. For this purpose we used classical MD with an interatomic potential, which was trained using a dataset of configuration, energies and forces, obtained from QMD calculations. 

\subsubsection*{Interatomic potential  \label{subsubsec:pot}}

At high pressures in the metal-hydrogen system, the contribution of many-body interactions to the energy is small with respect to the pair interaction as the bonding is predominantly metallic and the repulsion forces are dominant. Thus, only the contribution from the pair interactions to the energy was considered. This assumption is confirmed by the following validation of the potential. In this case: $E_{tot} = \frac{1}{2} \sum_{i,j=1}^{N_{at}} U(r_i, r_j)$. 

In fact, Cartesian coordinates cannot be used as descriptors of the atomic environment, because their numerical values are not invariant with respect
to translations and rotations of the system. Therefore, one should perform a preliminary procedure of \textit{symmetrization}. We used the method of \textit{symmetry functions} with functional form \eqref{eq:1}, which is a simplification of the approach proposed in \cite{li2015molecular, kruglov2017}.

\begin{equation}\label{eq:1} \tag{1}
X^{(k)}(r) = exp({-(\frac{r}{r_{cut, k}})^{p_k}}),
\end{equation}

where $p_k, r_{cut,k}$ are external parameters.Thus, one can take sums of the form \eqref{eq:2} as components of the feature vector which describes the local environment of the $i$-th atom.

\begin{equation} \label{eq:2} \tag{2}
X_{E,i}^{(k)} = \sum_{j=1}^{N_{neigh, i}} exp({-(\frac{|\vec{r_{i,j}}|}{r_{cut, k}})^{p_k}}),
\end{equation}

where $|\vec{r_{i,j}}|$ is the distance between atoms $i$ and $j$, $N_{neigh, i}$ is the number of nearest neighbors of the atom $i$ in the cutoff sphere of radius $R_{cut}$ (here $R_{cut} = 5.0$ \AA). 

Summing through all the atoms in the system, we obtain a symmetrized representation:

\begin{equation} \label{eq:3} \tag{3}
X_E^{(k)} = \sum_{i=1}^{N_{at}} \sum_{j=1}^{N_{neigh,i}} exp({-(\frac{|\vec{r_{i,j}}|}{r_{cut, k}})^{p_k}}), \quad \mathbf{X_E} = \{X_E^{(k)}\}_{k=1}^{k=N}.
\end{equation}

The obtained vectors $\mathbf{X_E}$ of length $N$ (where $k = 1, ... , N$, $N$ is the length of the $(r_{cut}, p)$ set) made up of such sums at different values $(r_{cut,k}, p_k)$ can be considered as features of the structures. Hence, we can present the energy of the system as a linear combination in the form \eqref{eq:4}:

\begin{equation} \label{eq:4} \tag{4}
E = \theta_0 + \mathbf{\Theta}^T \mathbf{X_E} 
\end{equation}
where $(\theta_0, \mathbf{\Theta})$ is the vector of regression coefficients. Therefore, the expression \eqref{eq:4} has a clear interpretation as a generalization of the Morse potential.

Using the presented method of symmetrization, it is possible to obtain a set of features for the forces on atoms by differentiation of expression \eqref{eq:2}:

\begin{align} \label{eq:5}
X_{F,i}^{l,(k)} = - \frac{\partial X_{E,i}^{(k)}}{\partial r_l} = \frac{\partial}{\partial r_l} \sum_{j=1}^{N_{neigh, i}} exp({-(\frac{|\vec{r_{i,j}}|}{r_{cut, k}})^{p_k}}) = \nonumber \\ = \sum_{j=1}^{N_{neigh, i}} exp({-(\frac{|\vec{r_{i,j}}|}{r_{cut, k}})^{p_k}}) (\frac{|\vec{r_{i,j}}|}{r_{cut, k}})^{p_k-1} \frac{p_k \ r_l}{r_{cut,k} \  |\vec{r_{i,j}}|} \tag{5}
\end{align}

Mathematical form of the expression \eqref{eq:5} allows us to construct a regression with the same coefficients $\mathbf{\Theta}$ for the vector $\mathbf{X_{F, i}^l}$ of length $N$. Thus, the $l$-th component of the force acting on the $i$-th atom in the system is given by expression \eqref{eq:6}:

\begin{equation} \label{eq:6} \tag{6}
F_i^l = \mathbf{\Theta}^T \mathbf{X_{F}}_i^l
\end{equation}

The approach just described is valid for single-component systems. In the case of two or more components, the interaction of different types of atoms is described by different coefficients $\mathbf{\Theta}$ and different sets of ($p, r_{cut}$). The expressions (\ref{eq:4}) and (\ref{eq:6}) for the two-component system with atoms of types A and B can be rewritten as follows:

\begin{align}\label{eq:7} 
E = \theta_0 + \mathbf{\Theta}^T_{A-A} \mathbf{X_{E}}^{A-A} + \mathbf{\Theta}^T_{A-B} \mathbf{X_{E}}^{A-B} + \nonumber \\ + \mathbf{\Theta}^T_{B-B} \mathbf{X_{E}}^{B-B}  \tag{7}
\end{align}
$$F_{A_i}^l = \mathbf{\Theta}^T_{A-A} \mathbf{X_F}_{A_i-A}^{l} + \mathbf{\Theta}^T_{A-B} \mathbf{X_F}_{A_i-B}^{l}$$
$$F_{B_i}^l = \mathbf{\Theta}^T_{A-B} \mathbf{X_F}_{B_i-A}^{l} + \mathbf{\Theta}^T_{B-B} \mathbf{X_F}_{B_i-B}^{l},
$$
where
\begin{equation}\label{eq:8} \tag{8}
X_{E}^{A-B, (k)}  = \sum_{i=1}^{N_{at}^A} \sum_{j=1}^{N_{neigh,A_i}^B} exp({-(\frac{|\vec{r_{i,j}}|}{r_{cut, k}^{A-B}})^{p_k^{A-B}}})
\end{equation}
\begin{align}\label{eq:9} 
X_{F,A_i-B}^{l, (k)}  = \sum_{j=1}^{N_{neigh, A_i}^B} exp({-(\frac{|\vec{r_{i,j}}|}{r_{cut, k}})^{p_k}}) (\frac{|\vec{r_{i,j}}|}{r_{cut, k}})^{p_k-1} \ \mathrm{x} \ \nonumber \\ \ \mathrm{x} \ \frac{p_k \ r_l}{r_{cut,k} \  |\vec{r_{i,j}}|}  \tag{9}
\end{align}

is a symmetrized representation for the energy and forces between atoms of types $A$ and $B$, and $N_{neigh, A_i}^B $ is the number of neighbors of type $B$ around $i$-th atom of type $A$. The representation for the other interactions can be obtained in the same way.

Constructed potential will yield an adequate prediction only if the interatomic distances will not differ much from those that were present in the training set. This becomes problematic $r \rightarrow 0 $, when the potential will turn to a constant. The absence of repulsion in the potential at small interatomic distances leads to the formation of dimers, which is an artifact. Thus, the potential has been modified to go to infinity at zero distance \eqref{eq:10}:

\begin{equation}\label{eq:10} \tag{10}
U(r) = U_0(r)  \theta(r-a) + (kr + b + (\frac{a}{r})^2 - 1) \theta(a-r)
\end{equation}

where $U_0 (r)$ is the potential obtained by linear regression, $\theta(r)$ is the Heaviside function and the coefficients $k, a$ provide smooth stitching of the potential. The starting point of extrapolation $a$ was selected from the condition of the minimum of potential function derivative.

\subsubsection*{Calculation parameters  \label{subsubsec:md}}

To carry out the analysis of the mixing process, several MD simulations were performed. In order to validate the quality of interatomic potential, the small system with 48 titanium atoms and 256 hydrogens was considered at the same initial conditions. The validation was made by the comparison of atomic structures (pure components and mixed system cases) and heat of mixing. After the validation, a larger calculation with the system of 2304 titanium atoms and 12144 hydrogen atoms in a unit cell of 10.8 x 10.8 x 425.8 \AA$^3$ with density of 4.1 g/cm$^3$ at the temperature of 4000 K was considered for calculations. Preliminary relaxation of the system was made using the gradient descent method and a subsequent low-temperature MD run at T = 300 K. Finally, MD simulations for 80 ps with a time step of 0.1 fs was conducted.

\section{Results and Discussion \label{sec:results}}

\subsection{Stable hydrides at P = 100 GPa \label{subsec:stable_hydrides}}

\begin{figure}[h!]
\centering
\includegraphics[width=0.45\textwidth]{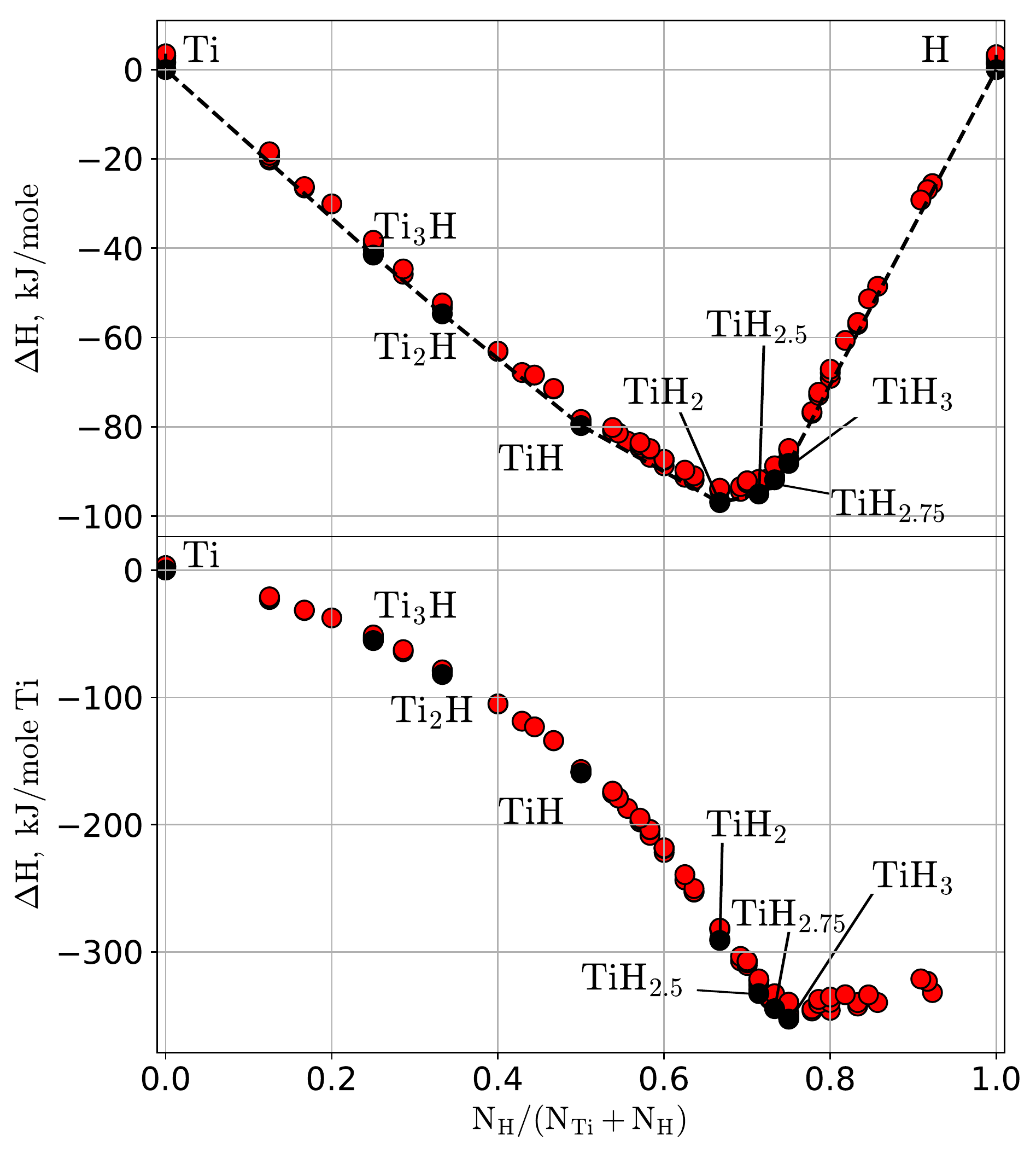}
\caption{Formation enthalpies of titanium hydrides  at $\mathrm{P=100} \ $ GPa in units of kJ/mole (top) and kJ/(mole Ti) (bottom).}
\label{fig:convex_hull}
\end{figure}

Using the evolutionary algorithm USPEX (Sec. \ref{subsec:uspex}) we predicted stable Ti-H phases with their enthalpies of formation ($\Delta H$). The obtained values of $\Delta H$ with corresponding compositions are shown in Figure \ref{fig:convex_hull}. 

The well-known titanium hydride is $\mathrm{TiH_2}$ \cite{stull1971, dantzer1976high, san1987h, zhang2006electron, wang1996thermodynamic, arita1982thermodynamics, china2008}, but we also predicted a number of stable phases with higher hydrogen content, namely: $\mathrm{TiH_{2.5}}$, $\mathrm{TiH_{2.75}}$, $\mathrm{TiH_{3}}$. One can note that the computational and experimental values of $\Delta H$ for $\mathrm{TiH_2}$ from \cite{stull1971, dantzer1976high, san1987h, zhang2006electron, wang1996thermodynamic, arita1982thermodynamics, china2008} vary from -123 to -179 kJ/(mole Ti), but at 100 GPa we find much more negative values: $\Delta H_{\mathrm{TiH_{2}}}$ = -290.88 kJ/(mole Ti) and $\Delta H_{\mathrm{TiH_{2.75}}}$ = -344.52 kJ/(mole Ti). Indeed, for titanium and many other metals the formation of hydrides becomes more preferable under pressure \cite{niobium, esfahani2016superconductivity, kruglov2017hs, kruglov2017uh, thorium, kvashnin2017iron}.

\subsection{Training and validation of the interatomic potential}

\begin{figure}[h!]
\center{
\includegraphics[width=0.47\textwidth]{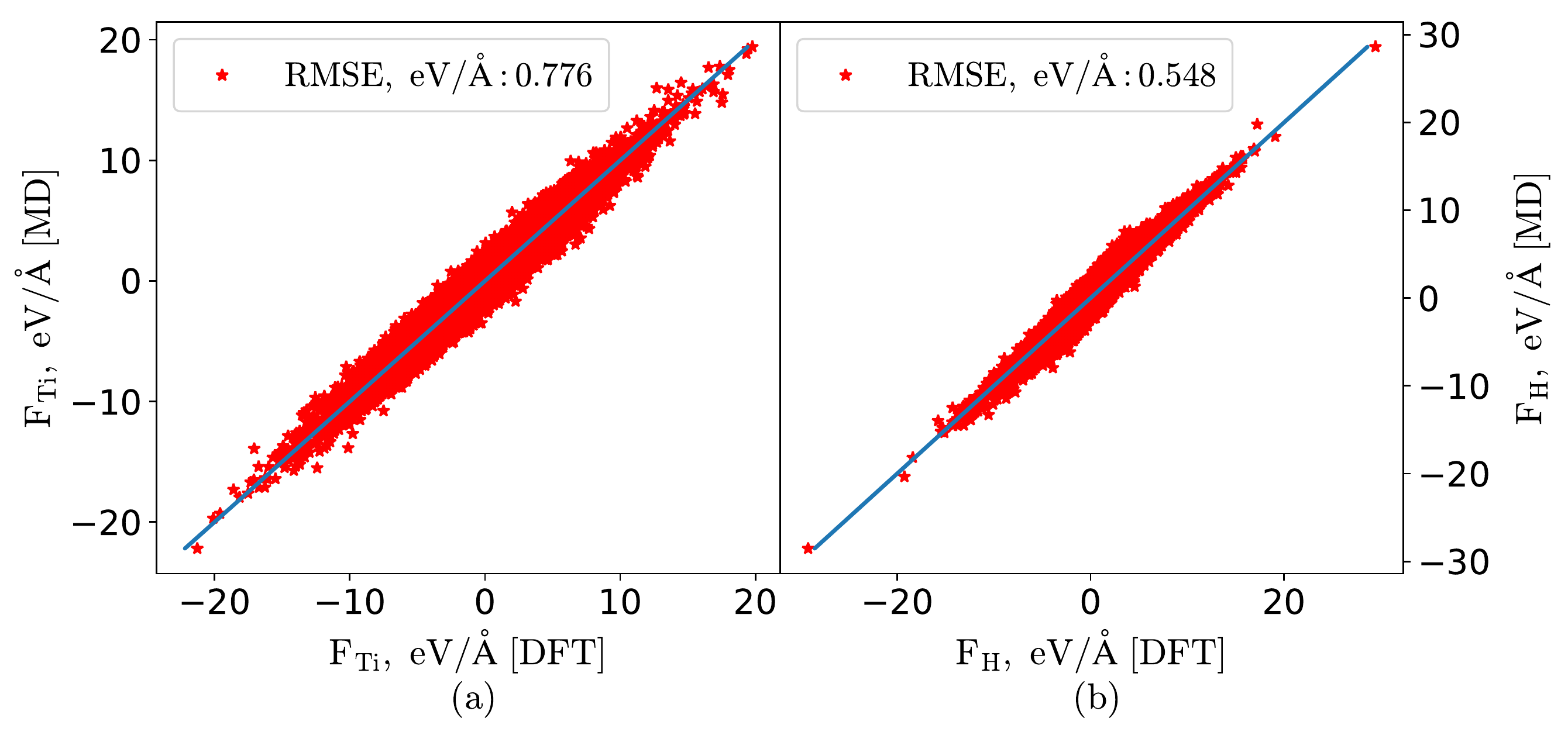}}
\caption{Comparison of ab-initio (x) and model (y) projections of forces acting on titanium atoms (a) and hydrogen atoms (b).}
\label{fig:errors}
\end{figure}

Previously we discussed the principles the interatomic potential is based on.  For this purpose one requires a set of features based on configurations and target values, which are typically interatomic forces and energies. We took expressions (\ref{eq:8} - \ref{eq:9}) as feature vectors, where the set of atomic coordinates as well as forces and energies of structures were obtained from the QMD calculations. A test of accuracy of this approach is shown in Figure \ref{fig:errors}. 

The values of RMSE were 0.776 eV/\AA \ and 0.548 eV/\AA \ for predictions of force projections on the titanium and hydrogen, respectively. Since the train and test data sets were based on NVE molecular dynamics, where total energy is conserved, here we make comparisons only for the forces. 

Fig. \ref{fig:errors} also shows that since all points lie along y=x line, the developed ML potential can be used to describe metal-hydrogen systems at discussed P-T conditions. 

\begin{figure}[h!]
\center{
\includegraphics[width=0.45\textwidth]{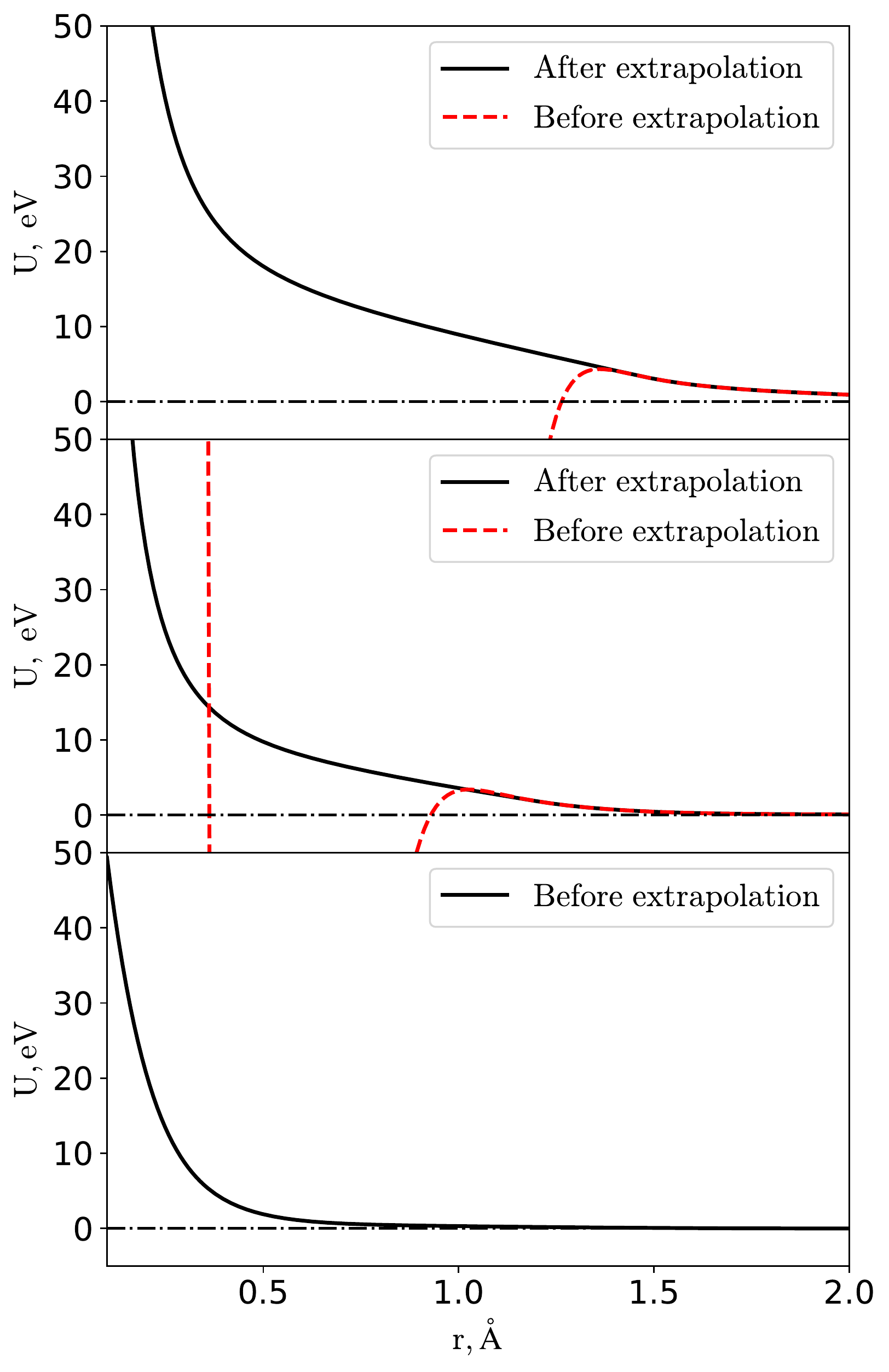}}
\caption{Dependence of the interaction energy $\mathrm{Ti-Ti}$ (top),  $\mathrm{Ti-H}$ (center) and  $\mathrm{H-H}$ (bottom) on the distance between atoms. Starting points for extrapolation: $a_{\mathrm{Ti-Ti}} = 1.44 $ \AA , $a_{\mathrm{Ti-H}} = 1.20  $ \AA. }
\label{fig:extr_plot}
\end{figure}

It should be noted that an extrapolation on the form of the potential was made at $r \rightarrow 0$. This extrapolation does not affect the potential at realistic distances, but removes pathologies displayed by uncorrected Ti-Ti and Ti-H potentials at extremely short distances (Figure \ref{fig:extr_plot}). 

This can be explained by the lack of data with short interatomic distances in the dataset. The starting points of the extrapolation (1.44 \AA \ for Ti-Ti potential, 1.2 \AA \ for Ti-H potential) are shorter than distances found in the RDF \ref{fig:rdf}. The role of our extrapolation correction is protection against random numerical errors and fluctuations.

\subsection{Atomic structures analysis \label{subsec:atom_analysis}}

\begin{figure}[h!]
\centering
\includegraphics[width=0.47\textwidth]{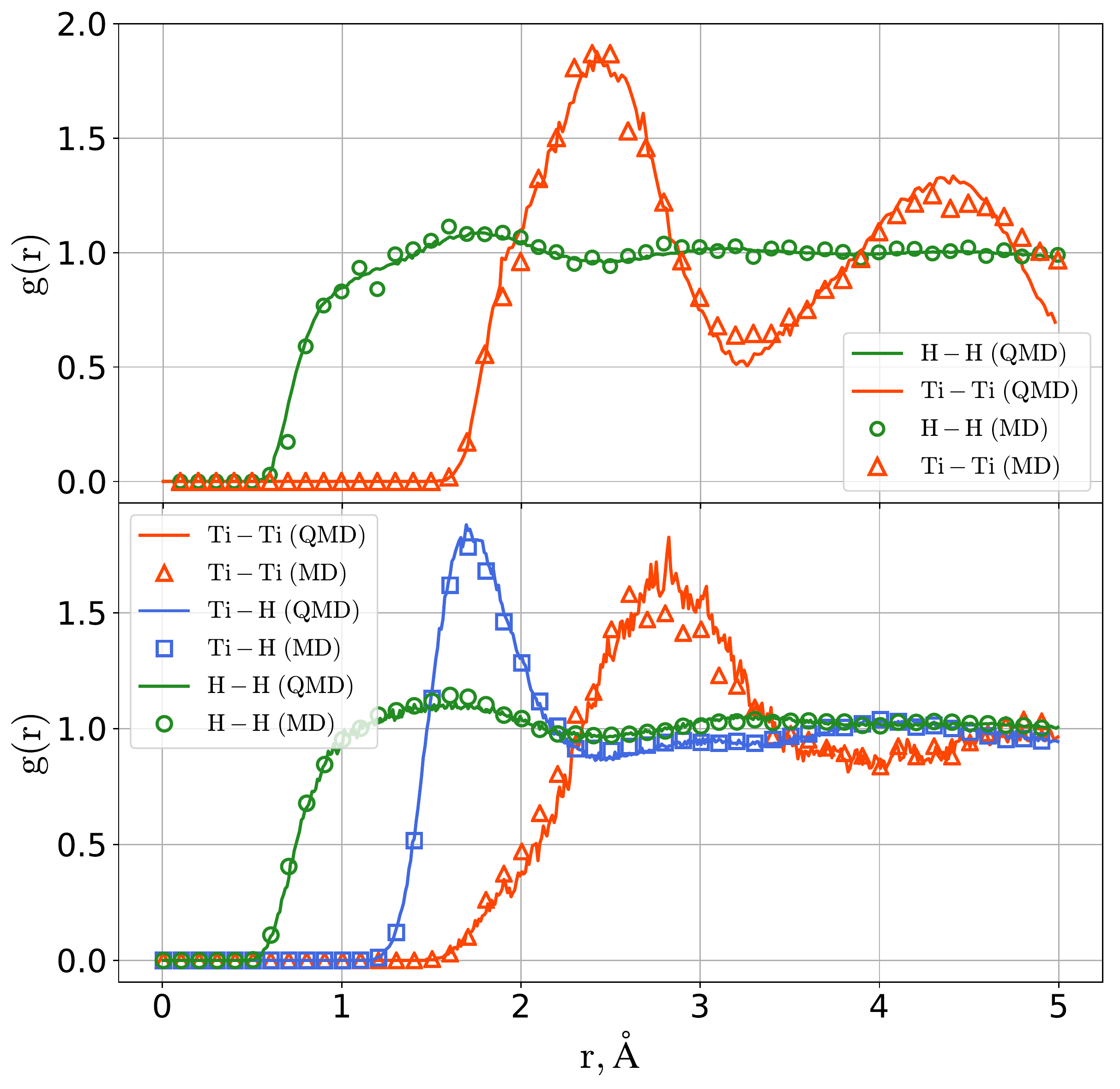}
\caption{RDF of the pure components (top, $\mathrm{T = 3000 \ K, \ \rho = 8.1 }$g/cm$^3$ for titanium and $\mathrm{T = 3600 \ K, \ \rho = 0.8 }$  g/cm$^3$ for hydrogen) and the mixture components (bottom, $\mathrm{T = 5600 \ K, \ \rho = 4.1 }$ g/cm$^3$) in MD and QMD calculations.}
\label{fig:rdf}
\end{figure}

Since the interatomic potential has to describe the mixing process of the Ti-H system, it should also provide the correct atomic structure of the pure components as well. We studied the atomic structures using RDF of the components in both these cases, making a comparison between the MD and QMD calculations of small systems. Calculations of pure elements were performed using titanium and hydrogen systems independently, where only corresponding parts of the interatomic potential were taken into account in the MD run. Analysis of the systems was carried out using the method discussed in Sec. \ref{subsec:QMD}, \ref{subsec:MD} (Fig. \ref{fig:rdf}). Values, obtained with the QMD calculation are plotted by solid line, while MD values are marked by symbols.

\begin{figure}[h!] 
\centering
\includegraphics[width=0.5\textwidth]{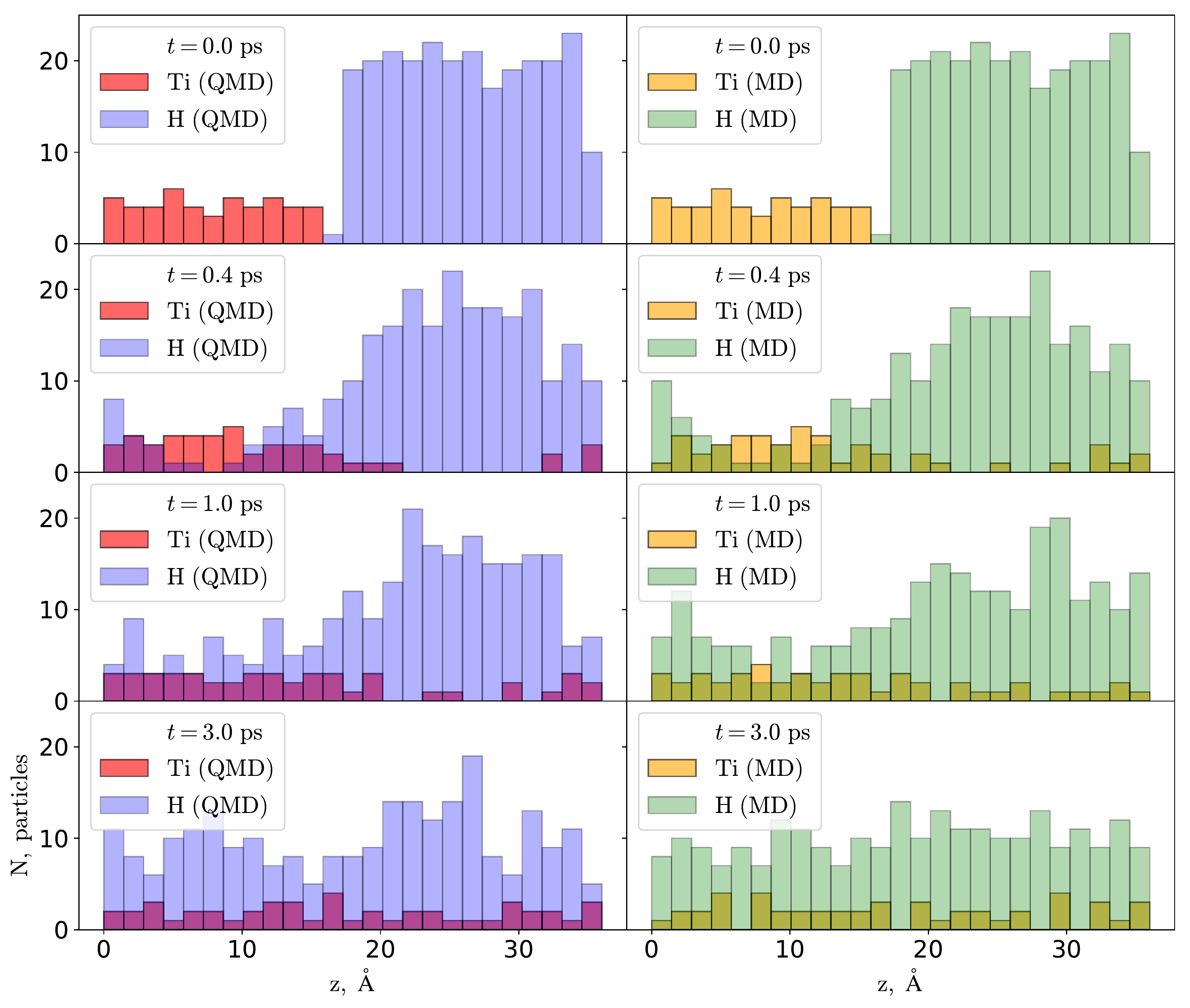}
\caption{Comparison of the components distribution profiles in MD and QMD calculations}
\label{fig:profiles}
\end{figure}

The results obtained by MD and QMD agree well with each other, which proves the correctness of our ML potential. Good agreement between characteristic peaks, total behavior of RDF for pure hydrogen and reference data from \cite{norman2017} (the comparison was made with T = 4000 K and $\rho = $ 0.9 g/cm$^3$) was observed. Therefore, the comparison of mixing in the discussed cases was performed from equal initial conditions. Distribution profiles of the components along the selected direction (z-axis) were compared at certain points of time: 0.0, 0.4, 1.0, and 3.0 ps respectively (Figure \ref{fig:profiles}). 

\begin{figure*}[!htbp]
\centering
\includegraphics[width=0.7\textwidth]{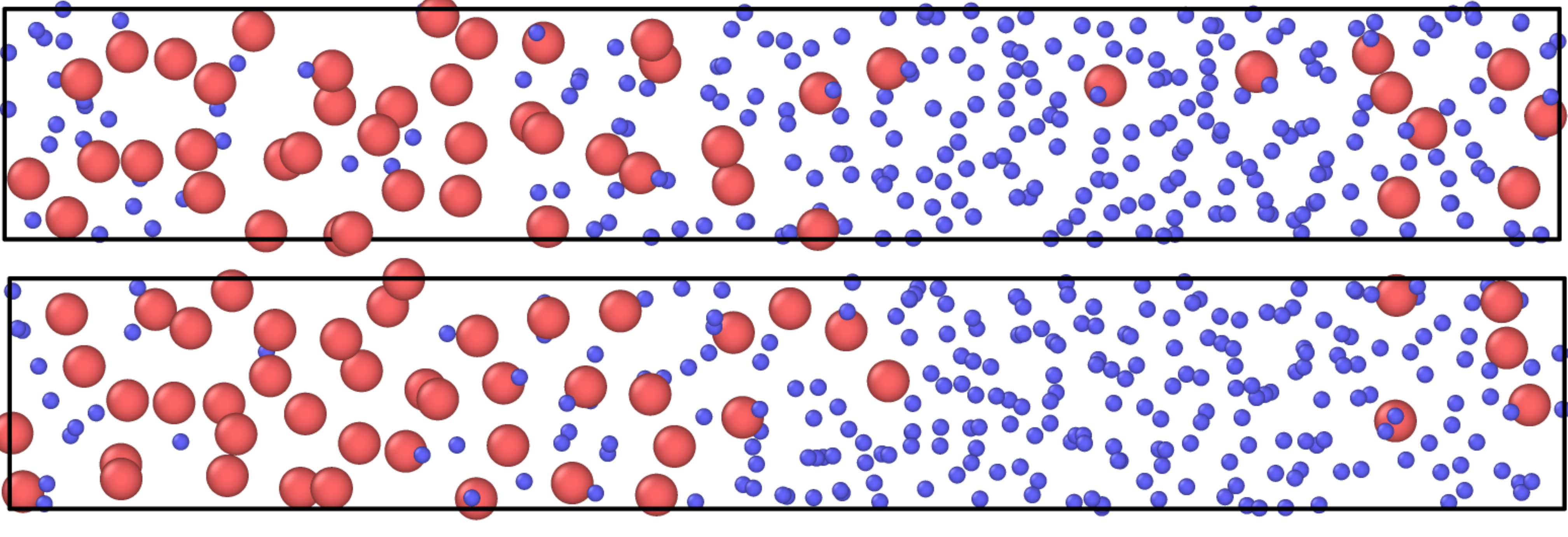}
\caption{Visualization of $\mathrm{Ti-H}$ system during mixing in MD (top) and QMD (bottom) calculations, t = 0.4 ps.}
\label{fig:visual}
\end{figure*}

From the Fig. \ref{fig:profiles}, one can see that in both (MD and QMD) calculations the active process of mutually diffusive penetration of atoms begins immediately.Titanium becomes saturated with hydrogen at t = 1.0 ps. Visualization of the system during mixing is shown in Figure \ref{fig:visual}. At t = 3.0 ps the system appears to be completely mixed up.

\subsection{Heat effect of mixing \label{subsec:heat}}

\begin{figure}[h!]
\centering
\includegraphics[width=0.45\textwidth]{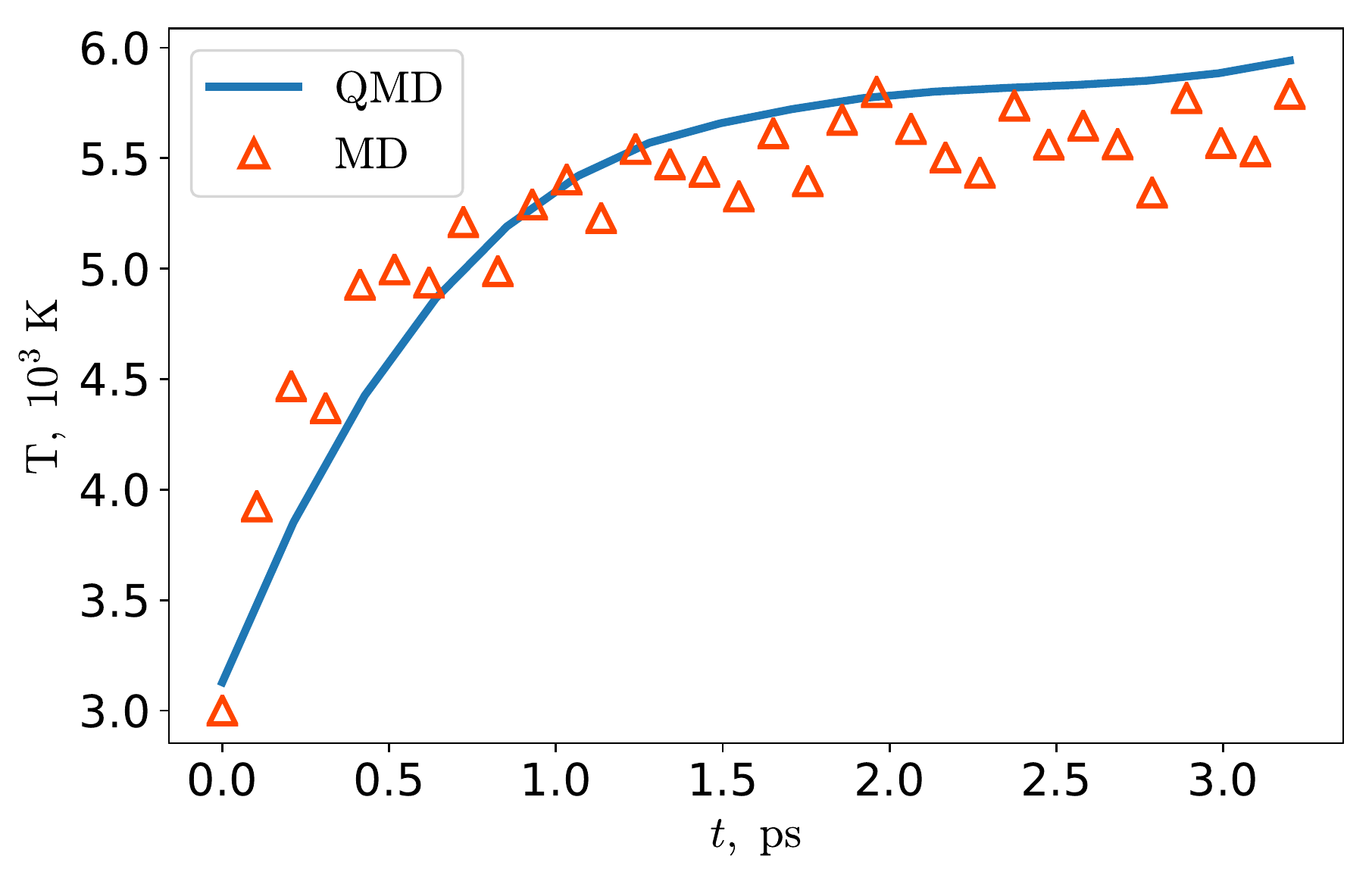}
\caption{Dependence of the mixture temperature on time during the mixing process}
\label{fig:entalpy}
\end{figure}

One of the main thermodynamic characteristics of the mixing process is the enthalpy of mixing, which determines the heat effect of the process. It allows one to determine the character of the reaction between mixing fluids. Employing the temperature difference between initial and final states, we calculated the mixing enthalpy using formula \eqref{eq:11} (at constant pressure).

\begin{equation} \tag{11} \label{eq:11}
    \Delta H =  c_{\mu}  \frac{N_{Ti} + N_{H}}{N_{Ti}} \Delta T
\end{equation}

where $ c_{\mu}$ is molar specific heat, $\Delta T$ is the temperature difference, $\mathrm{N_{Ti}}, \mathrm{N_{H}}$ is numbers of titanium and hydrogen atoms, and $R = 8.31 $ J/(mole $\cdot$ K). The value of specific heat for the given temperature range was evaluated in MD run and is equal to $ c_{\mu} \simeq 2.33 \ R$. This result is consistent with the phonon theory of liquids \cite{bolmatov2012}. The thermodynamics of mixing was analyzed using the calculations described in Sec. \ref{subsec:QMD}, \ref{subsec:MD}. The dependence of temperature of the mixture on time in MD and QMD cases is given in Figure \ref{fig:entalpy}. 

Since the temperature of the system becomes constant at t = 1.0 ps, further dissolution does not result in any heat effect. The fraction of absorbed hydrogen was estimated using Fig. \ref{fig:profiles} at t = 1.0 ps by taking the ratio of the number of hydrogen atoms penetrated into the titanium layer to the number of titanium atoms in this layer: $x = N_{H} / N_{Ti} \simeq 2.7$. The obtained value of enthalpy of mixing at 100 GPa is $\Delta H_{\mathrm{TiH_{2.7}}} = $ -327.75 kJ/(mole Ti). This result is in robust agreement with enthalpy of formation of  $\mathrm{TiH_{2}}$ and $\mathrm{TiH_{2.75}}$ hydrides, obtained with the USPEX method: $\Delta H_{\mathrm{TiH_{2}}}$ = -290.88 kJ/(mole Ti) and $\Delta H_{\mathrm{TiH_{2.75}}}$ = -344.52 kJ/(mole Ti)  (Sec. \ref{subsec:stable_hydrides}). 

\subsection{Mixing character}

\begin{figure}[h!]
\centering
\includegraphics[width=0.45\textwidth]{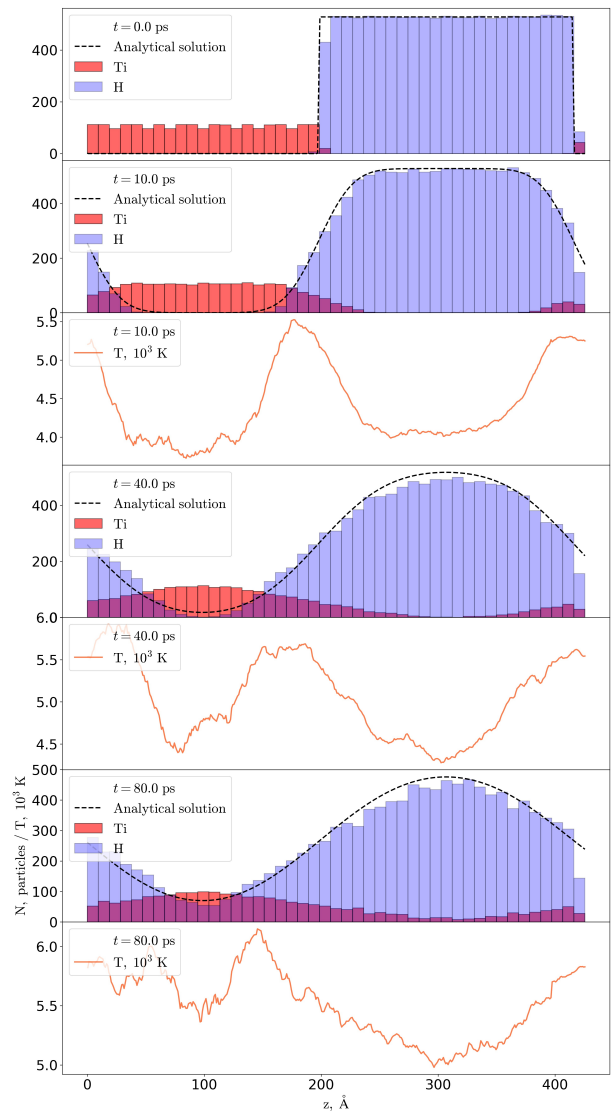}
\caption{Temperature and concentration profiles of mixture components during the MD calculation. Analytical predictions of hydrogen concentration are plotted by a dotted line.}
\label{fig:prof_temp}
\end{figure}

The performed calculations show that the constructed interatomic potential is reliable and allow us to carry out simulations of large systems. In this section, we considered the mixing process from the position of classical diffusion theory, employing the method discussed in Sec. \ref{subsec:MD}. Histograms in Figure \ref{fig:prof_temp} show the particle distribution profiles in the MD calculation with the developed MD potential (the distribution is given along the z-axis as in Fig. \ref{fig:profiles}). 

We compared hydrogen penetration dynamics (histogram in Fig. \ref{fig:prof_temp}) with the analytical solution of the  one-dimensional diffusion equation \eqref{eq:12} (dotted line). The diffusion coefficient $D$ in \eqref{eq:12} (calculated using  Einstein law \eqref{eq:13} in the equilibrium mixture at T = 4000 K) is equal to 27.7 \AA$^2$/ps. 

\begin{equation} \label{eq:12} \tag{12}
\frac{\partial c}{\partial t} = \frac{\partial^2}{\partial x^2}(D c) 
\end{equation}

\begin{equation} \label{eq:13} \tag{13}
\langle \vec{r}^2 \rangle = 6Dt 
\end{equation}

Strong agreement between the analytical curve and hydrogen distribution profile indicates that the mixing process at the discussed region of the phase diagram has diffusive character. In the same figure one can see the temperature profiles along the z-axis. There is a notable increase in temperature observed in the regions of exothermic mixing, reaching a maximum at the boundary between the interacting components.

\subsection{Estimation of the dissolution time of ejecta}

\begin{figure}[h!]
\centering
\includegraphics[width=0.49\textwidth]{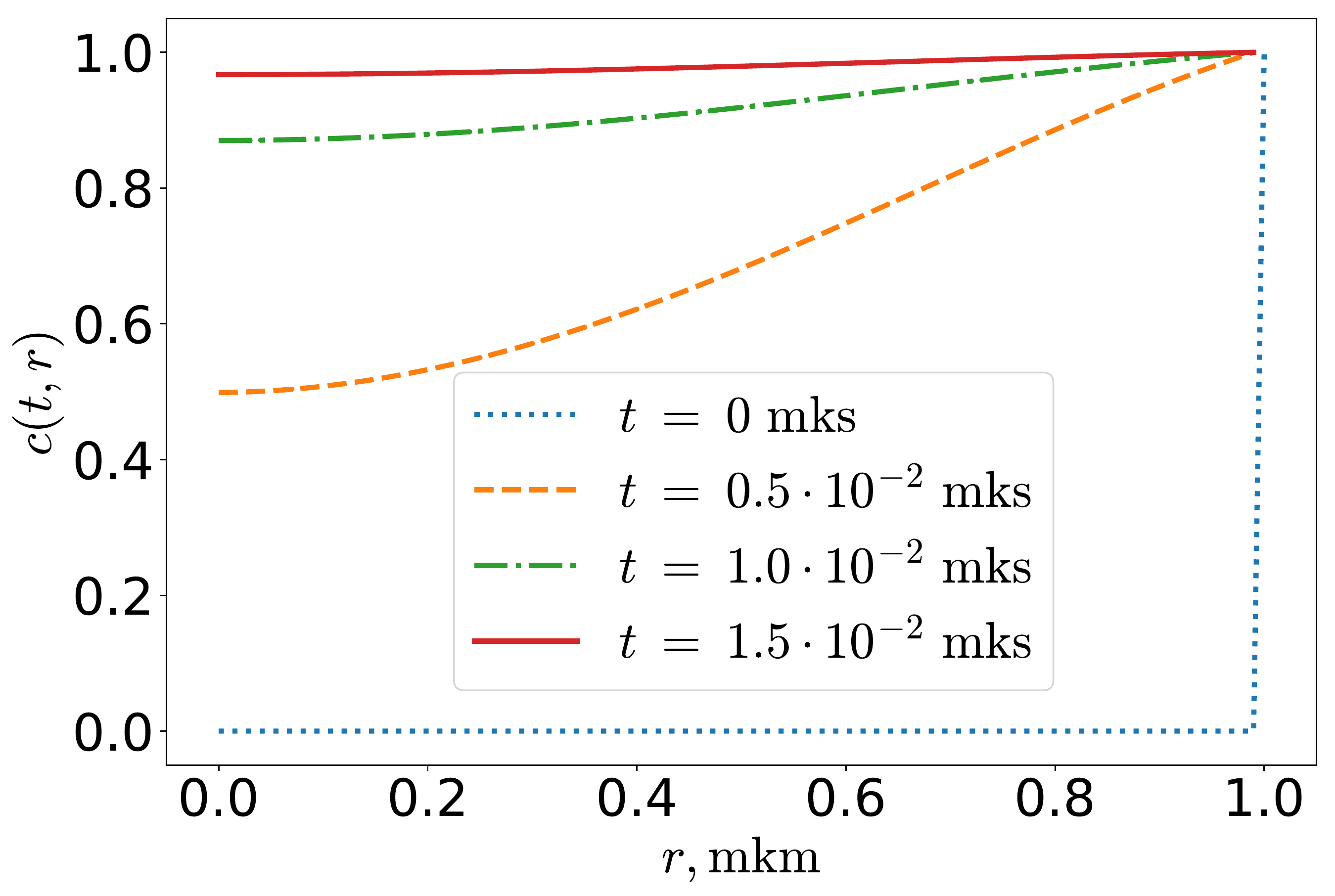}
\caption{Hydrogen concentration profiles from 3D diffusion equation.}
\label{fig:eqn3D}
\end{figure}

Knowledge of diffusive character of the mixing process allows to make an estimation of the ejecta particles' dissolution time. For this purpose, a spherical titanium particle with a radius beyond 1 $\mu m$ was considered. Evaluation was made using an analytical solution of the 3D diffusion equation (Figure \ref{fig:eqn3D}). At the starting point (t = 0), hydrogen atoms (given in $\mathrm{c / c_{max}}$ units) are only at the region of r = 1 $\mu m$, which represents the edge of the sphere. Therefore, the diffusion process of hydrogen atoms inside the ejecta dissolves the particle within a duration of 1.5$\cdot 10^{-2}$ $\mu s$.

\section{Conclusion} 
In this work, we studied the process dissolution of titanium ejecta particles in warm dense hydrogen at megabar pressure. The presented approach shows that the process can be described by diffusion law at T $> 3 \cdot 10^3$ K and megabar pressure. Furthermore, the limit of saturation for titanium was absent i.e. all available titanium in the system was completely dissolved in hydrogen, occupying the entire volume of the computational cell. The final state of the Ti-H system was found as a homogenized fluid with completely dissolved titanium particles. In the case of ejecta particles with a radius of 1 $\mu m$, we found that complete dissolution occurs within  $1.5 \cdot 10^{-2}$ $\mu s$. This result can be generalized to a wider range of pressures and temperatures where titanium and hydrogen are atomic fluids.

\section{Acknowledgments}

We thank O. V. Sergeev for integration of interatomic potential in LAMMPS code and I. A. Kruglov for the useful discussions. The calculations were performed on a cluster of the Dukhov Research Institute of  Automatics.

\bibliographystyle{unsrt}
\bibliography{references}

\end{document}